# ELT Contributions to The First Explosions[1]

A Whitepaper Submitted to the Astro 2020 Decadal Survey Committee

J. Craig Wheeler (The University of Texas at Austin)
József Vinkó (Konkoly Observatory)
Rafaella Margutti (Northwestern University)
Dan Milisavljevic (Purdue University)
Maryam Modjaz (New York University)
Sung-Chul Yoon (Seoul National University)

Contact information for Primary Author:
J. Craig Wheeler
Department of Astronomy, The University of Texas at Austin
2515 Speedway, C1400
Austin, TX 78712-1205
512-471-6407
wheel@astro.as.utexas.edu

---

[1] [1] Adapted from a chapter in the 2018 edition of the Science Book of the Giant Magellan Telescope Project.

**The large aperture and sensitive optical and near infrared imager spectrographs will enable an ELT system to observe some supernovae at large distances, deep into cosmological history when supernovae first began to occur.**

After the Big Bang, the Universe became dark. The cosmic dark ages ended with the formation of the first stars at z ~20, only ~200 Myr after the Big Bang. These stars started cosmic reionization, created the first heavy elements, and illuminated the first galaxies. Some of these early stars may have created the first black holes. No foreseeable facility will be able to directly detect those first stars, but some of them will explode to produce bright, detectable sources that will yield insight into the rate of production and nature of those first stars. The first explosions will also point the way to deep studies of the host environment. As we discuss below, a wide range of different kinds of supernovae may be visible in the first generations of stars.

It has long been recognized that long, soft gamma-ray bursts should be detectable at very high redshifts. Because their continua rise to the blue, redshifting brings more flux into the observer frame and compensates for the increased distance. If the first stars produced GRBs, then their study can provide information on the nature of the explosion to complement the burgeoning body of evidence from relatively nearby events at redshifts less than 6 to 8. GRBs could be seen in principle at redshifts of 30 or more and track the literal onset of the end of the dark ages. The line-of-sight spectra of these events will probe the intergalactic media through which their flux propagates.

Another potentially critical probe of the first stars are superluminous supernovae (SLSN). Hydrogen-deficient SLSN I, in particular, are commonly observed in low-metallicity, star-forming galaxies. This makes them especially promising events to track the rate of star formation at high redshift and to seek any evolution of the intrinsic properties of the explosion at the earliest era. SLSN I are especially bright in the UV, a property that will aide their detectability at high redshift. Their observed rate per unit volume increases at least to z ~ 4, beyond the peak in the star formation rate at z ~ 2 to 3 (Cooke et al. 2012). The large UV luminosity can last for months in the rest frame and correspondingly longer by 1 + z at higher redshift. LSST will have detection limits fainter than $m_{AB}$ ~ 24 mag; in principle LSST can see SLSN up to z ~ 3 (Figure 1). The redshift evolution of SLSN can provide evidence for the physical origin of SLSN and for how the stellar initial mass function changes with redshift.

As an example of the power of an ELT system, we illustrate the potential of the proposed *GMACS* low-dispersion spectrograph on the *Giant Magellan Telescope*. Analogous arguments could be made for instruments on the *Thirty Meter Telescope*.

The *GMT* could produce a 5σ seeing-limited spectrum in an hour at 0.5 μm at $m_{AB}$ = 25. (*GMACS* will achieve a S/N ratio of about 10 in an hour-long exposure at a resolution of R = 2000 down to about 24$^{th}$ magnitude. The performance of *NIRMOS* is expected to be comparable). For an SLSN I with M ~ -22 this corresponds to z ~ 3 (within the detection limits of the *LSST*, as discussed above). For imaging, the limit at 0.5 μm drops to $m_{AB}$ = 27 at a redshift of z ~ 4. For observations at 2 μm, the seeing-limited spectroscopic limit is $m_{AB} \sim 24$ at $z \sim 2$. With AO at 2 μm, the spectroscopic limit is $m_{AB} \sim 27$ corresponding to $z \sim 10$. With AO at 2 μm, the imaging limit of $m_{AB} \sim 28$ substantially exceeds $z = 10$. SLSN will be sparse at these redshifts. Special techniques utilizing the deep imaging capabilities of an ELT system may be required to detect them. SLSN clearly represent a powerful probe of the early Universe, and an ELT system can see them.

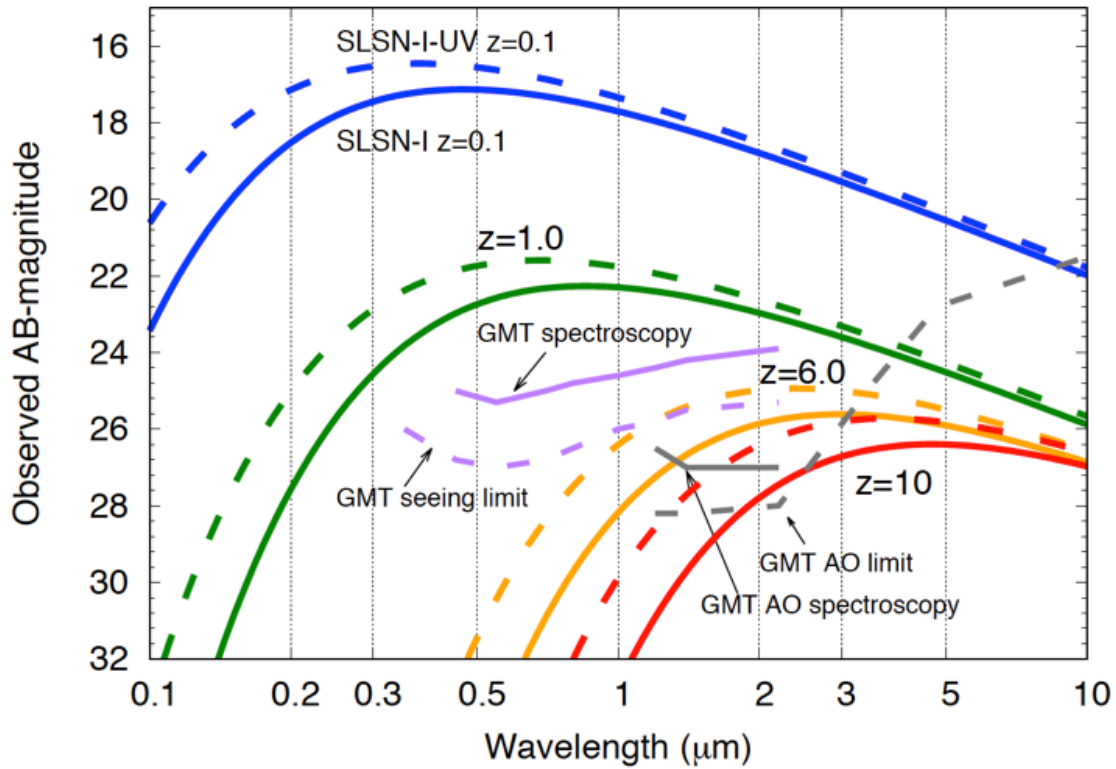

*Figure 1. The GMT is expected to detect superluminous supernovae (SLSN), among the first stars that are born and explode at the end of the cosmic dark ages. The figure presents the predicted AB magnitude of SLSN I as a function of wavelength at a variety of redshifts (Figure courtesy of J. Vinkó). Solid lines correspond to a photospheric temperature of 12,000 K, dashed lines to 15,000 K. With AO at 2 μm, spectra could be obtained at redshift 10 and imaging could reach to even higher redshifts. GMACS can get spectra of SLSN to redshift 3 at 0.5 μm. Seeing-limited imaging could extend to redshift 6.*

Theoretical models suggest the possibility of pair-instability supernovae (PISN) arising in massive stars that are hot enough to create de-stabilizing electron/positron pairs (Barkat et al. 1967). These could also be visible at high redshifts. Such models predict thermonuclear explosion of their oxygen cores and total disruption of the star. Models predict that some PISN may produce a large amount of radioactive $^{56}$Ni that could yield exceptionally bright events, depending on the mass and whether a hydrogen envelope were retained. Some of these could be seen at large redshift, if rarely (Hummel et al. 2012). Due to time dilation, PISN are expected to be detectable for about 10 years in the observer frame.

The first Type Ia supernovae (SN Ia) must await the sedate processes of lower mass stars in binary systems. The first SN Ia are expected at z ~ 3 to 5. At that redshift, SN Ia will be within 1 mag of peak light for about 100 days in the local frame. At 0.5 μm, SN Ia will drop out at z ~1 because of the strong UV deficit (Figure 2). For an SN Ia with M ~ -19, the GMT could produce a 5σ seeing-limited spectrum in an hour at 2 μm to somewhat less than z = 1. With AO at 2 μm, the spectroscopic limit is $m_{AB}$ ~ 27 corresponding to z ~ 3. With AO at 2 μm, the imaging limit of $m_{AB}$ ~ 28 extends to z ~ 6. Beyond z ~ 6, SN Ia would again drop out because of their UV deficiency as well as the delay time needed for the formation of their progenitor systems. In principle, the GMT could directly observe the epoch of turn-on of SN Ia. This would bring a revolution in our understanding of galaxy evolution, star formation history, stellar evolution in both single and binary systems and the progenitors and explosion mechanism of SN Ia. A critical question is whether the first SN Ia resemble "typical" SN Ia on which supernova cosmology is based, or are biased in some way.

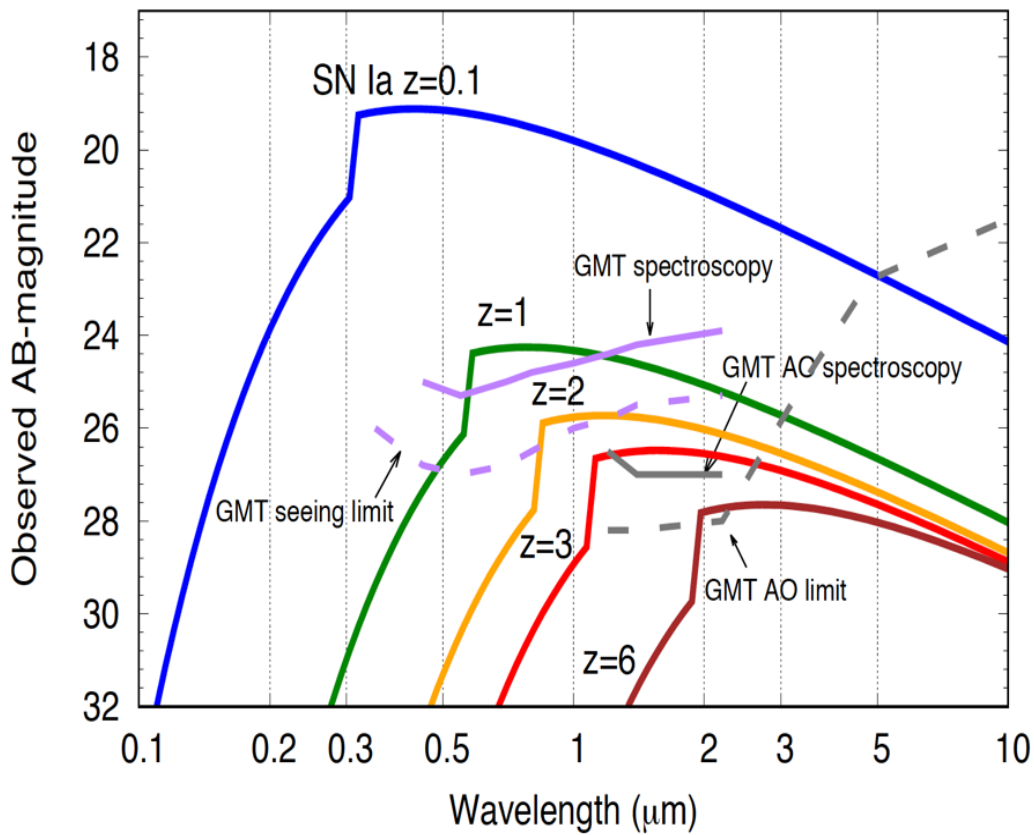

*Figure 2 The stellar populations in galaxies at redshifts 3-5 are the right age to produce SN Ia from low mass stars in binary systems, and the GMT is capable of probing this critical era. The figure presents the predicted AB magnitude of SN Ia as a function of wavelength at a variety of redshifts (figure courtesy of J. Vinkó). Note the steep drop of flux to the blue at each epoch. GMTIFS can get spectra of SN Ia to z~1 at 2 μm. With AO, spectra could be obtained at z~3. AO imaging could extend to z~6.*